\documentclass[graybox]{svmult}

\usepackage{type1cm}        
\usepackage{makeidx}         
\usepackage{graphicx}        
\usepackage{multicol}        
\usepackage[bottom]{footmisc}

\usepackage{newtxtext}       %
\usepackage{newtxmath}       

\makeindex             

\usepackage{url}
\usepackage{xspace}
\usepackage{wrapfig}
\usepackage{amsmath}
\usepackage{pifont}
\usepackage{tikz}
\usetikzlibrary{calc}
\usetikzlibrary{positioning}
\usetikzlibrary{automata}
\usetikzlibrary{shapes}
\usetikzlibrary{decorations.pathreplacing}
\usepackage{pgfplots}

\usepackage{listings}
\newcommand{\qchar}[1]{\mathsf{{}'#1'}}

\makeatletter
\let\UrlSpecialsOld\UrlSpecials
\def\UrlSpecials{\UrlSpecialsOld\do\/{\Url@slash}\do\_{\Url@underscore}}%
\def\Url@slash{\@ifnextchar/{\kern-.11em\mathchar47\kern-.2em}%
    {\kern-.0em\mathchar47\kern-.08em\penalty\UrlBigBreakPenalty}}
\def\Url@underscore{\nfss@text{\leavevmode \kern.06em\vbox{\hrule\@width.3em}}}
\makeatother

\begin{document}

\title*{JBMC}
\subtitle{A Bounded Model Checking Tool for Java Bytecode\thanks{The work in this paper is partially funded by the EPSRC grants EP/T026995/1, EP/V000497/1, and EU H2020 ELEGANT 957286.}}

\author{Romain Brenguier, Lucas Cordeiro, Daniel Kroening and Peter Schrammel}
\institute{
Romain Brenguier \at Diffblue Ltd, United Kingdom \email{romain.brenguier@diffblue.com}
\and Lucas C. Cordeiro \at University of Manchester, United Kingdom \email{lucas.cordeiro@manchester.ac.uk}
\and Daniel Kroening \at University of Oxford, United Kingdom and Diffblue Ltd, United Kingdom \email{kroening@cs.ox.ac.uk}
\and Peter Schrammel \at University of Sussex, Brighton, United Kingdom and Diffblue Ltd, United Kingdom \email{p.schrammel@sussex.ac.uk}}

\maketitle

\begin{abstract}
JBMC is an open-source SAT- and SMT-based bounded model checking tool for verifying Java bytecode. JBMC relies on an operational model of the Java libraries, which conservatively approximates their semantics, to verify assertion violations, array out-of-bounds, unintended arithmetic overflows, and other kinds of functional and runtime errors in Java bytecode. JBMC can be used to either falsify properties or prove program correctness if an upper bound on the depth of the state-space is known. Practical applications of JBMC include but are not limited to bug finding, property checking, test input generation, detection of security vulnerabilities, and program synthesis. Here we provide a detailed description of JBMC's architecture and its functionalities, including an in-depth discussion of its background theories and underlying technologies, including a state-of-the-art string solver to ensure safety and security of Java bytecode.
\end{abstract}

\section{Introduction} 
\label{sec:introduction}

The Java Programming Language is a general-purpose, concurrent, strongly typed, object-oriented language~\cite{javadoc2017}. Applications written in Java are compiled to the bytecode instruction set and binary format as defined in the Java Virtual Machine (JVM) specification.\footnote{\url{https://docs.oracle.com/javase/specs/index.html}} This compiled Java bytecode can run on all platforms
on top of a JVM without the need for recompilation. However, Java programs may have bugs, which may result in array out-of-bounds, unintended arithmetic overflows, and other kinds of functional and runtime errors. In~addition, Java allows multi-threading, and thus, problems such as race conditions and deadlocks can occur.

There exist various verifiers for Java programs (or its bytecode instruction representation) available in literature that employ explicit-state, symbolic model checking or static checking techniques~\cite{DBLP:conf/tacas/AnandPV07,DBLP:journals/sigplan/FlanaganLLNSS13,DBLP:conf/cav/KahsaiRSS16}. Here we describe a Bounded Model Checking Tool for Verifying Java Bytecode named JBMC, which is a sophisticated verification tool for programs compiled to Java Bytecode. Its development started in $2015$ on top of the CProver framework~\cite{CKL04}. JBMC employs Bounded Model Checking (BMC)~\cite{handbook09} in combination with Boolean
Satisfiability (SAT) and Satisfiability Modulo Theories (SMT)~\cite{BarrettSST09} and full symbolic state-space exploration, which allows developers to perform a bit-accurate verification of Java programs. It primarily aims to help Java developers by finding subtle bugs in their code. In particular, it allows verifying user-specified assertions, runtime exceptions (e.g., \textsf{ArrayOutOfBoundsException}), and exceptions that escape the entry point and thus cause the JVM to abort. Note that JBMC does not require any individual annotations in the Java code to find such bugs. However, users can specify their own assertions to verify custom properties.

JBMC can be used to solve various common verification problems such as bug finding, property checking, test input generation, detection of security vulnerabilities, and program synthesis. These verification problems can arise from different domains, such as embedded systems, web and enterprise applications. JBMC is open-source (under a BSD-style license), and its source code
is available at \url{https://github.com/diffblue/cbmc/tree/develop/jbmc}.

\section{Verification Approach} 
\label{sec:approach}


JBMC is an extension to the C Bounded Model Checker
(CBMC)~\cite{DBLP:conf/tacas/ClarkeKL04}, named
JBMC,
that verifies Java bytecode.  JBMC consists of a frontend for
parsing Java bytecode and a Java operational model (JOM), which is an
exact but verification-friendly model of the standard Java libraries.
In JBMC, the Java bytecode to be analyzed is modelled as a state transition
system, which is extracted from the control-flow
graph (CFG)~\cite{Muchnick97}. This graph is built as part of a
translation process from the Java bytecode to static single assignment (SSA)
form. A node in the CFG represents either a (non-) deterministic 
assignment or a conditional statement, while an edge
in the CFG represents a possible change in the bytecode's control
location.  

Given the transition system \textit{M}, a property $\phi$, and a bound
\textit{k}, JBMC unrolls the system \textit{k} times and translates it
into a verification condition that is satisfiable \textit{iff} $\phi$
has a counterexample of length $k$ or less.
The verification condition is a quantifier-free formula in a decidable
subset of first-order logic, which is then checked for satisfiability
by a SAT solver.\footnote{JBMC natively supports MiniSAT as its main
  solver to discharge VCs and check for their satisfiability, but can
  also be used with other incremental SAT solvers such as Glucose.}
In particular, JBMC's backend produces formulae using the
\texttt{QF\_ABVFP+Strings} logic, where \texttt{QF} stands for
quantifier-free formulas, \texttt{A} stands for the theory of arrays,
\texttt{BV} stands for the theory of fixed-sized bit-vectors and
\texttt{FP} stands for the theory of
floating-points~\cite{Barrett10c.:the}.

More formally, the associated model checking problem 
in JBMC is given by the following logical formula:
\begin{equation} 
  \forall s_0,\ldots,s_k:
I(s_{0}) 
\wedge 
  \bigwedge^{k-1}_{i=0} T(s_{i},s_{i+1})
\Longrightarrow 
  \bigwedge^{k}_{i=0}\phi(s_i)
\end{equation}

Here, $\phi$ is a safety property, $I$ the set of 
initial states of $M$ and $T(s_{i},s_{i+1})$ the 
transition relation of $M$ between time steps $i$ and $i+1$. Hence,
$I(s_0)\wedge\bigwedge^{k-1}_{i=0} T(s_{i},s_{i+1})$ represents the
executions of $M$ of length $k$.
The formula is valid \textit{iff}
$\phi_{i}$ holds for all steps up to $k$.

Instead of checking validity of the formula,
we negate it and check for satisfiability:

\begin{equation} 
  \label{eq:bounded-model-checking}
I(s_{0}) 
\wedge 
  \bigwedge^{k-1}_{i=0} T(s_{i},s_{i+1})
\wedge 
  \bigvee^{k}_{i=0}\neg\phi(s_i)
\end{equation}

Equation~\eqref{eq:bounded-model-checking} can be satisfied
\textit{iff} for some $i \leq k$ there exists a reachable state at
time step $i$ in which $\phi$ is violated.  If
Equation~\eqref{eq:bounded-model-checking} is satisfiable, then $\phi$
is violated and the SAT solver provides a satisfying assignment, from
which we can extract the values of the program variables to construct
a counterexample. A counterexample for a property $\phi$ is a sequence
of states $s_{0}, s_{1},\ldots, s_{k}$ with $s_{0} \in S_{0}$, $s_{k}
\in S$, and $T\left(s_{i}, s_{i+1}\right)$ for $0 \leq i < k$. If
Equation~\eqref{eq:bounded-model-checking} is unsatisfiable, we can
conclude that no error state is reachable in $k$ steps or less.

A distinct feature of JBMC's verification approach, when compared with other
approaches~\cite{DBLP:conf/tacas/AnandPV07,DBLP:journals/sigplan/FlanaganLLNSS13,DBLP:conf/cav/KahsaiRSS16},
is the use of Bounded Model Checking (BMC)~\cite{handbook09} in
combination with Boolean Satisfiability and Satisfiability Modulo
Theories (SMT)~\cite{BarrettSST09} and full symbolic state-space
exploration, which allows us to perform a bit-accurate verification of
Java programs based on \texttt{QF\_ABVFP+Strings} logic.

\section{Software Project}
\label{sec:project}

JBMC is based on the CPROVER framework.%
\footnote{\url{https://www.cprover.org}}
JBMC is implemented in C++ and currently has around 20 KLOC
(not counting CPROVER itself).
The source code is available on \url{https://www.github.com/diffblue/cbmc}
in the \textsf{jbmc} directory under a BSD license.
JBMC is easy to build from source on many platforms
including Linux, Mac OS and Windows.

A Java library called CPROVER Java API (\textsf{cprover-api.jar},
also BSD license)
provides the interface between JBMC and Java programs.%
\footnote{\url{https://github.com/diffblue/java-cprover-api}}
The CPROVER Java API is also available in the Maven central repository.%
\footnote{\url{https://search.maven.org/artifact/org.cprover.util/cprover-api}}

The Java operational model (JOM) uses the CPROVER Java API
to access JBMC functionality such as non-deterministic assignment
or \textsf{assume}.%
\footnote{\url{https://github.com/diffblue/java-models-library}}
The JOM (\textsf{core-models.jar}) is based on OpenJDK 8 and is
therefore provided separately under GPL2 with classpath exception.

Table~\ref{tab:features} gives an overview of JBMC's features.

\begin{table}[t]
  \begin{tabular}{|p{0.25\textwidth}|p{0.7\textwidth}|}
    \hline
    Languages  & Java Bytecode \\
    \hline
    Properties & \textsf{assert}, runtime exceptions, uncaught exceptions \\
    \hline
    Environments  & Linux, Mac OS, Windows \\
    \hline
    Technologies used  & symbolic execution, bounded model checking \\
                       & SAT and SMT solving \\
    Current strengths  & strings, floating point \\
    \hline
    Current weaknesses & multi-threading, Java library support \\
    \hline
  \end{tabular}
  \caption{JBMC Features}
  \label{tab:features}
\end{table}

\section{Use Cases} 
\label{sec:usecases}


JBMC can be used to verify array bound violations, unintended arithmetic
overflows, and other kinds of functional and runtime errors in Java bytecode.
Runtime errors in Java (e.g., illegal memory access) are detected by the JVM
and an appropriate exception is thrown (e.g., \texttt{NullPointerException},
\texttt{ArrayIndex\-OutOfBoundsException}).  An \texttt{AssertionError} is
thrown on violation of a condition specified by the programmer using the
\texttt{assert} keyword.  JBMC analyzes the program and verifies whether
such error conditions occur.

JBMC's command line options are similar to those of the \texttt{java}
executable.
JBMC can be used to analyze a class file:\footnote{%
If a class \textsf{MyClass} is in a package \texttt{my.package}, then compile
it to \textit{some-dir}\texttt{/my/package/MyClass.class}, and in \textit{some-dir} execute
\textit{jbmc-installation-dir}\texttt{/jbmc my.package.MyClass {-}-unwind }$k$.}

\vspace*{0.5ex}
\begin{verbatim}
jbmc MyClass --unwind k
\end{verbatim}
\vspace*{0.5ex}

\noindent or a Java archive (\texttt{.jar}) file:  

\vspace*{0.5ex}
\begin{verbatim}
jbmc -jar file.jar --main-class MyClass --unwind k
\end{verbatim}
\vspace*{0.5ex}

In both cases the entry point for the analysis of the program is the
\texttt{public static void main} method of the specified main class.
If the given Java archive has no main class attribute in its manifest
then the main class can be specified using the \texttt{{-}-main-class}
option.
The classpath can be given using the \texttt{-cp} option
as customary when running a program with the \texttt{java} executable.

$k$ is a
positive integer limiting the number of times loops are unwound and
recursions are unfolded.
If no bug is found, up to a $k$-depth unwinding, then JBMC reports
\texttt{VERIFICATION SUCCESSFUL}; otherwise, it reports \texttt{VERIFICATION
  FAILED} along with a counterexample in the form of an execution
trace (\texttt{{-}-trace}), which contains the full variable
assignment on each program state with file, method, and line
information. 

Note that if the Java bytecode is compiled with debug information
(\texttt{javac -g} option), then JBMC can also provide the original
program variable names in the counterexample, rather than just
bytecode variable slots.  Further options in JBMC can be retrieved via
\texttt{jbmc} \texttt{{-}-help}.

The examples mentioned in the following sections can be found at
\url{https://github.com/diffblue/cbmc/tree/develop/jbmc/regression/book-examples}.

%
%



\begin{figure}[t]
\centering
  \begin{lstlisting}[language=java,escapechar=^]
public class StringUtil {
  public static String getLastToken(
      String toSplit, char delimiter, int limit) {
    if(toSplit == null) {
      return null;
    }
    int numberOfTokens = 0;
    int tokenStart = 0;
    int tokenEnd = toSplit.indexOf(delimiter, tokenStart);
    while (tokenEnd != -1 && numberOfTokens < limit - 1) {
      ++numberOfTokens;
      tokenStart = tokenEnd + 1;
      tokenEnd = toSplit.indexOf(delimiter, tokenStart);
    }
    ++numberOfTokens;
    String lastToken = toSplit.substring(tokenStart, toSplit.length());
    assert(lastToken.indexOf(delimiter) < 0 || numberOfTokens == limit);^\label{line:functional-example:assert}^
    return lastToken;
  }
}
\end{lstlisting}
\caption{Example of a functional property}
\label{fig:functional-example}
\end{figure}

\begin{figure}[t]
\centering
\begin{lstlisting}[language=HTML,escapechar=^]
...
Results:
...
[StringUtil.getLastToken.assertion.1]
  assertion at file StringUtil.java line 17
  function StringUtil.getLastToken bytecode-index 52:
  FAILURE
...
VERIFICATION FAILED  
\end{lstlisting}
\caption{Results for the example of a functional property in Figure~\ref{fig:functional-example}.}
\label{fig:functional-example-results}
\end{figure}

\begin{figure}[t]
\centering
\begin{lstlisting}[language=HTML,escapechar=^]
...
Trace for StringUtil.getLastToken.assertion.1:
...
StringUtil.java line 4 function __CPROVER__start
  delimiter='\ua775'

StringUtil.java line 4 function __CPROVER__start
  limit=0
...
StringUtil.java line 14 function StringUtil.getLastToken
  numberOfTokens=1
...
\end{lstlisting}
\caption{Trace for the example of a functional property in Figure~\ref{fig:functional-example}.}
\label{fig:functional-example-trace}
\end{figure}

\subsection{Functional Properties}

Functional properties can be specified in Java programs
with the help of \texttt{assert} statements.
Let us consider \texttt{StringUtil.java} shown in
Figure~\ref{fig:functional-example}.
The method \texttt{getLastToken} contains such an \texttt{assert}
statement in line~\ref{line:functional-example:assert}.
It is expected that when splitting the string given by
\texttt{toSplit} using \texttt{delimiter} into a maximum number
of \texttt{limit} tokens then the \texttt{lastToken} does not
contain a delimiter if we have not reached the limit.
We attempt to verify this property by executing the following command:%
\footnote{On Linux and Mac OS systems. On Windows the classpath elements
need to be separated by \texttt{;} instead of \texttt{:}.}
\begin{verbatim}
jbmc StringUtil.getLastToken
  --classpath cprover-api.jar:core-models.jar:.
  --max-nondet-string-length 100 --unwind 2
\end{verbatim}
\noindent where \texttt{--classpath} adds the CPROVER API
(\texttt{cprover-api.jar}) and the JOM
(\texttt{core-models.jar}) to the classpath;
\texttt{{-}-max-nondet-string-length} bounds the maximum length of
non-de\-ter\-min\-istic strings.
Note that the entry point is specified to be a method other than
the \texttt{main} method here, namely \texttt{getLastToken}.

For this particular example, JBMC reports the results
as shown in Figure~\ref{fig:functional-example-results}.
This means that the assertion does not hold.

We can add the \texttt{{-}-trace} option to understand why.
JBMC then displays the trace as shown in
Figure~\ref{fig:functional-example-trace}.
This shows that our implementation does not behave according to
specification in the case where \texttt{limit} equals \texttt{0}.
That is likely because we have not even thought what should happen
for non-positive limits. So, we have to specify this case and
implement accordingly.

\begin{figure} [t]
\centering
\begin{minipage}{8.5cm}
  \begin{lstlisting}[language=java,escapechar=^]
public class BinarySearch {
  public static int binarySearch(int[] array, int value) {
    int lowerBound = 0;
    int upperBound = array.length;^\label{line:exception1}^
    int i = array.length / 2;
    while(array[i] != value && upperBound > lowerBound + 1) {^\label{line:exception2}^
      if(array[i] > value) upperBound = i; else lowerBound = i;
      i = (upperBound + lowerBound) / 2;
    }
    return i;
  }
}
\end{lstlisting}
\end{minipage}
\caption{Example of detecting uncaught runtime exceptions}
\label{fig:exception-example}
\end{figure}

\begin{figure}[t]
\centering
\begin{lstlisting}[language=HTML,escapechar=^]
...
Results:
...
[BinarySearch.binarySearch:([II)I.1] no uncaught exception: FAILURE
...
VERIFICATION FAILED  
\end{lstlisting}
\caption{Results for the runtime exception example in Figure~\ref{fig:exception-example}.}
\label{fig:exception-example-results}
\end{figure}

\begin{figure}[t]
\centering
\begin{lstlisting}[language=HTML,escapechar=^]
...
Trace for BinarySearch.binarySearch:([II)I.1:
...
file BinarySearch.java line 3 function __CPROVER__start
  array=null
...
file BinarySearch.java line 4 function BinarySearch.binarySearch
  dynamic_object4.@class_identifier="java::java.lang.NullPointerException"
...
\end{lstlisting}
\caption{Trace for the runtime exception example in Figure~\ref{fig:exception-example}.}
\label{fig:exception-example-trace}
\end{figure}

\subsection{Runtime Exceptions}


Runtime Exceptions in Java derive from the \textit{RuntimeException}
class, which can be thrown during the normal operation of the
JVM~\cite{javadoc2017}. Runtime exceptions are not declared in a
method's throws clause.
Figure~\ref{fig:exception-example} shows \texttt{BinarySearch.java},
an implementation of the binary search algorithm to find the index of
an integer in an array.
We can run JBMC to automatically check for uncaught runtime exceptions
in this program by executing the following command:
\begin{verbatim}
jbmc BinarySearch.binarySearch
  --throw-runtime-exceptions --unwind 2
\end{verbatim}
\noindent where \texttt{{-}-throw-runtime-exceptions} makes implicit
runtime exceptions explicit and detects them when they \emph{escape}
the entry point method.
Without \texttt{{-}-throw- runtime-exceptions} JBMC would report a failed
property when an exception is thrown even though it is caught by
a surrounding \texttt{try-catch}.

Running the above command JBMC reports the results shown in
Figure~\ref{fig:exception-example-results}.
It tells us that the property \texttt{no uncaught exception} failed,
i.e.\ there is an uncaught exception.
We can find out which type of exception has escaped by running with
the \texttt{{-}-trace} option.

JBMC shows the counterexample trace in
Figure~\ref{fig:exception-example-trace}.
A \texttt{NullPointerException} is thrown in
line~\ref{line:exception1} if the \texttt{array} argument is
\texttt{null}.
If we tried to fix this problem and check again then we will notice
that the method can throw a further runtime exception, an
\texttt{ArrayIndexOutOfBoundsException} in line~\ref{line:exception2},
when it is called with an empty array.

\subsection{Test Inputs}

\begin{figure}[t]
\centering
\begin{lstlisting}[language=java,escapechar=^]
public class LocatorHandler {
  public enum LocatorType {XPATH, ID; }
  public static Locator autoLocator(String locator, LocatorType type) {
    switch (type) {
    case XPATH:
      if (locator.startsWith("xpath=")) {
        // assert(false);
        locator = locator.substring(locator.indexOf("=") + 1);^\label{test-inputs:cover1}^
      }
      return new XPathLocator(locator);
    case ID:
      if (locator.startsWith("id=")) {
        // assert(false);
        locator = locator.substring(locator.indexOf("=") + 1);^\label{test-inputs:cover2}^
      }
      return new IdLocator(locator);
    }
    return null;
  }
}
\end{lstlisting}
\caption{Example for creating test inputs}
\label{fig:test-inputs}
\end{figure}

\begin{figure}[t]
\centering
\begin{lstlisting}[language=HTML,escapechar=^]
Results:
...
[LocatorHandler.autoLocator.assertion.1] 
  assertion at file LocatorHandler.java line 7
  function LocatorHandler.autoLocator bytecode-index 14:
  FAILURE
  
[LocatorHandler.autoLocator.assertion.2] 
  assertion at file LocatorHandler.java line 13
  function LocatorHandler.autoLocator bytecode-index 37:
  FAILURE
...
\end{lstlisting}
\caption{Results for the test inputs example in Figure~\ref{fig:test-inputs}.}
\label{fig:test-inputs-results}
\end{figure}

\begin{figure}[t]
\centering
\begin{lstlisting}[language=HTML,escapechar=^]
Trace for LocatorHandler.autoLocator.assertion.1:
...
LocatorHandler.java line 4 function __CPROVER__start
  dynamic_object2={ 'x', 'p', 'a', 't', 'h', '=', 'a' }
...
LocatorHandler.java line 4 function __CPROVER__start
  dynamic_object1.data=dynamic_object2
...
LocatorHandler.java line 4 function __CPROVER__start
  INPUT locator: &dynamic_object1
...
java/lang/Enum.java line 118 function java.lang.Enum(java.lang.String, int)
  dynamic_object3.@java.lang.Enum.name=&XPATH
...
LocatorHandler.java line 4 thread 0 function __CPROVER__start
  INPUT type: &dynamic_object3
...
\end{lstlisting}
\caption{Trace for the test inputs example in Figure~\ref{fig:test-inputs}.}
\label{fig:test-inputs-trace}
\end{figure}

For the class in Figure~\ref{fig:test-inputs} we would like to write
tests covering certain branches in the source code.  To achieve this,
we need to come up with inputs, reaching
lines~\ref{test-inputs:cover1} and~\ref{test-inputs:cover2}, for
instance.  This can be done by adding \texttt{assert(false);}
instructions before these instructions and running JBMC with the
\texttt{{-}-trace} option:
\begin{verbatim}
jbmc LocatorHandler.autoLocator
  --classpath cprover-api.jar:core-models.jar:.
  --max-nondet-string-length 10 --unwind 10 --trace
\end{verbatim}
In the output shown in Figure~\ref{fig:test-inputs-results} we see
that both program locations have been reached, i.e.\ the corresponding
assertions failed.

We can then inspect the trace for how the inputs are initialized. For
example, for the assertion in line~7, we see the values shown in
Figure~\ref{fig:test-inputs-trace}.  This means that invoking the
method with \texttt{locator="xpath=a"} and \texttt{type=XPATH} will
cover the statement in line~\ref{test-inputs:cover1}.
Similarly, we can find the inputs required to reach
line~\ref{test-inputs:cover2}.

\subsection{Security Vulnerabilities}

JBMC can be used to find potential security flaws, for instance by
identifying untrusted strings that are handled to sensitive functions
without being first sanitized.  In the example of
Figure~\ref{fig:security}, if the \texttt{xml} string comes from an
untrusted source, it can be harmful to the system executing this code
as an attacker could inject arbitrary Java code into the program.

\begin{figure}[t]
\centering
\begin{lstlisting}[language=java,escapechar=^]
import java.io.StringReader;
import javax.xml.bind.JAXBContext;
import javax.xml.bind.JAXBException;
import javax.xml.bind.Unmarshaller;
import javax.xml.stream.XMLInputFactory;
import javax.xml.stream.XMLStreamException;
import javax.xml.stream.XMLStreamReader;
import org.w3c.dom.Comment;

public class SimpleXXE {
  protected Comment parseXml(String xml)
      throws JAXBException, XMLStreamException {
    JAXBContext jc = JAXBContext.newInstance(Comment.class);
    XMLInputFactory xif = XMLInputFactory.newFactory();
    xif.setProperty(XMLInputFactory.IS_SUPPORTING_EXTERNAL_ENTITIES, true);
    xif.setProperty(XMLInputFactory.IS_VALIDATING, false);
    xif.setProperty(XMLInputFactory.SUPPORT_DTD, true);
    XMLStreamReader xsr = xif.createXMLStreamReader(new StringReader(xml));
    Unmarshaller unmarshaller = jc.createUnmarshaller();
    return (Comment) unmarshaller.unmarshal(xsr);
  }
}
\end{lstlisting}
\caption{Illustrative Java code for a security flaw}
\label{fig:security}
\end{figure}

To detect these cases using JBMC, we could instrument the
code by declaring a \texttt{InstrumentedString} class with a Boolean
flag \texttt{is\_tainted} which is set to \texttt{false} for strings
coming directly the program but \texttt{true} when coming from the
user.  Then an \texttt{assert(!arg.is\_tainted);} assertion is added
in sensitive functions. It is possible to have sanitizing functions
which set the flag to \texttt{false}.
To perform such instrumentations automatically, a taint analysis could
be used.  The CPROVER framework has a tool called \texttt{JANALYZER}
which has limited experimental support for taint analysis.

\subsection{Equivalence Checking}

For large software systems, the verification effort to completely
re-check the entire software from scratch might be too high, and is
largely wasted if, as is often the case, the changes are
small~\cite{OHearn18}. Continuous software verification plays an
important role to determine which parts of a given software
system need to be re-verified by obtaining information from previous
verification runs~\cite{ChudnovCCDHMMMM18}.  In this respect, JBMC can
be used to prove that two versions of a given method are
computationally equivalent.

\begin{figure}[t]
\centering
\begin{minipage}{0.7\textwidth}
\begin{lstlisting}
public class SignalUtil {
  public static long abs1(int signal) {
    long result;
    if(signal >= 0)
      result = signal;
    else
      result = -1*signal;
    return result;
  }
  public static long abs2(int signal) {
    if(signal < 0)
      return -signal;
    else
      return signal;
  }
}
\end{lstlisting}
\end{minipage}
\centerline{(a)}
\\[2.5ex]
\begin{minipage}{0.7\textwidth}
\begin{lstlisting}
public class EquivalenceCheck {
  public void check(int signal) {
    assert(SignalUtil.abs1(signal) == SignalUtil.abs2(signal));
  }
}
\end{lstlisting}
\end{minipage}
\centerline{(b)}
\caption{(a) Original function \texttt{abs1} and optimized version
  \texttt{abs2}. (b) Harness for equivalence checking.}
\label{fig:equivalence-checking} 
\end{figure}

As an example, consider the two versions of a function for computing
the absolute value of a signal, illustrated in
Figure~\ref{fig:equivalence-checking}(a).  They
were extracted from the software of two releases of a medical device
product~\cite{CordeiroFCM09}.  In order to prove the equivalence of
these two methods, we compare their input-output relations.

We thus write an equivalence checking harness as illustrated in
Figure~\ref{fig:equivalence-checking}(b).
This harness checks that the two versions behave equivalently
if they produce the same outputs provided that they are supplied
with the same inputs.
Inputs are any data read by the compared programs and outputs are any
data modified by the programs, including return values and exceptions
that they throw. Care must be taken in writing the harness if the
programs have side effects on static fields or modify the input
arguments.
In concrete terms for our simple example, the returned values must match
if the inputs are the same.
We can then check by running

\begin{verbatim}
jbmc EquivalenceCheck.check
\end{verbatim}

\noindent and obtain the result shown in
Figure~\ref{fig:equivalence-checking-results}, proving that the two
versions are input/output-equivalent.

\begin{figure}[t]
\centering
\begin{lstlisting}[language=HTML,escapechar=^]
Results:
...
[EquivalenceCheck.check:(I)V.assertion.1]
  assertion at file EquivalenceCheck.java line 3
  function EquivalenceCheck.check:(I)V bytecode-index 11:
  SUCCESS
...
\end{lstlisting}
\caption{Results for equivalence checking the example in Figure~\ref{fig:equivalence-checking}.}
\label{fig:equivalence-checking-results}
\end{figure}
\section{Architecture} 
\label{sec:architecture}

\begin{figure}[t]
\centering
\includegraphics[width=1.0\textwidth]{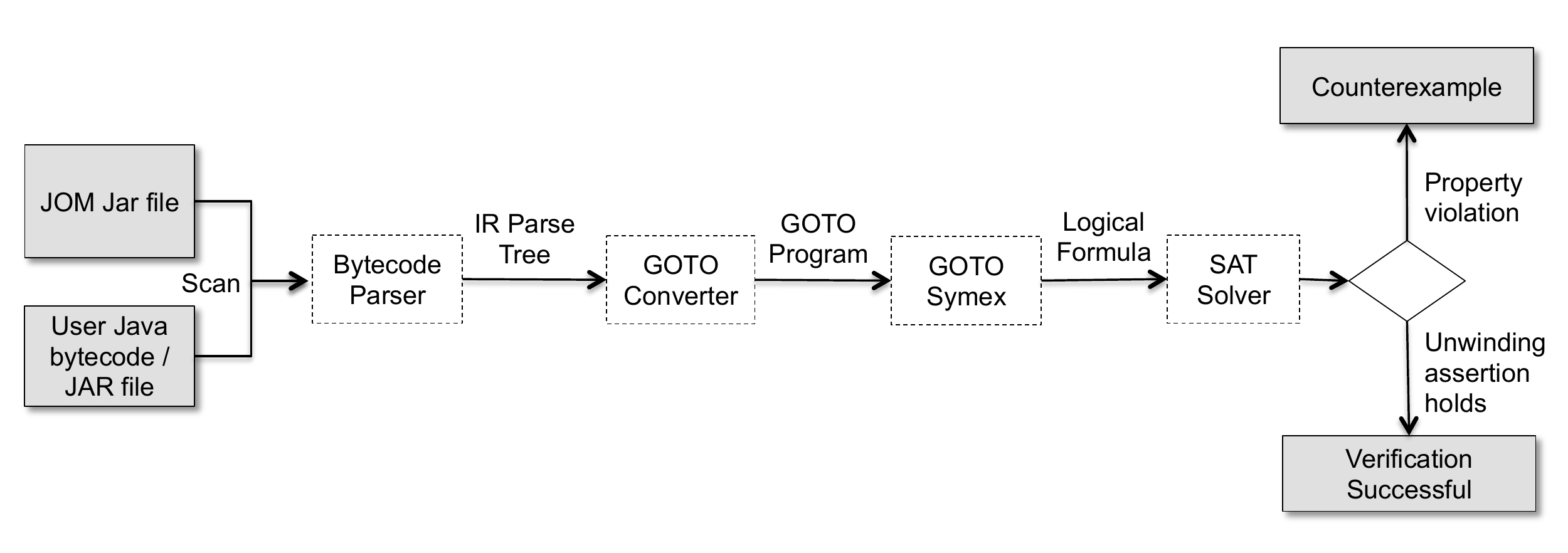}
\caption{JBMC Architecture~\cite{CKKST18}. Grey rectangles represent input and output while white rectangles represent the JBMC main verification steps.}
\label{figure:jbmc-arch}
\end{figure}

JBMC's architecture (see Figure~\ref{figure:jbmc-arch}) is a pipeline
of passes that transform the input Java bytecode from one format to
another until the verification results are obtained.

\subsection{Front-End}

Our front-end integrates a class loader, which accepts Java bytecode
\emph{class} files and \emph{jar} archives.  The parse trees for the classes
are translated into the CPROVER CFG representation, which is called a
\emph{GOTO} program~\cite{DBLP:conf/tacas/ClarkeKL04}.

\subsection{GOTO}

The GOTO program component converts the Java bytecode into a goto-program,
which simplifies the representation (e.g., replacement of \textit{switch} and \textit{while} by 
\textit{if} and \textit{goto} statements). The GOTO symex component performs a symbolic simulation 
of the Java bytecode, which thus handles the unrolling of the loops and the elimination of recursive
functions; and generates the verification conditions to be encoded in the SAT/SMT back-end.


\subsection{Lowerings}




\subsubsection{Polymorphism}

Polymorphism allows variable instances to be bound to references of different types, related by inheritance~\cite{Alexander02}. JBMC handles polymorphism by encoding virtual method dispatch into a \textit{switch} statement over the runtime type information attached to the object to select the correct method to be called. In particular, when a Java class declares non-static methods, which are by default ``virtual methods'', JBMC creates a virtual table, which contains a pointer to each virtual method in the class. If a derived class does not override a virtual method, then the pointers are copied to the derived class's virtual table. In contrast, if a derived class overrides a virtual method, then the pointers in the derived class's virtual table will point to the overridden method implementation. Whenever a virtual method is called, JBMC executes the method pointed in the virtual table via a switch statement using ITE operators from the SMT-lib~\cite{Barrett10c.:the}.

\subsubsection{Exception Handling}

Exceptions are unexpected circumstances that arise during the execution of a program, \textup{e.g.}, runtime errors~\cite{Deitel2019}. In Java, the exception handling is split into three (basic) elements: a \textit{try} block, where a thrown exception can be directed to a \textit{catch} statement; a set of \textit{catch} statements, where a thrown exception can be handled; and a \textit{throw} statement that raises an exception. Similarly to polymorphism, JBMC encodes the complex control flow arising from exceptions into conditional branches. We record the exception thrown in a global variable, which is then used to propagate the exception up the call stack until a matching \textit{catch} statement (if any) to handle the error is reached. To support exception handling in JBMC, we extended our GOTO conversion code and the symbolic engine. In the former, we had to define new instructions and model the throw expression as jumps. In the latter, we implemented the rules for throwing and catching exceptions. JBMC can detect when the JVM would abort due to an exception that is not caught within the program.

\subsection{Symbolic Execution}


The resulting \emph{GOTO} program is then passed to the bounded model
checking algorithm for finding bugs.  The BMC algorithm symbolically
executes the program, unwinding loops and unfolding recursive function calls
up to a given bound.  The resulting bit-vector formula is then passed on to
the configured SAT or SMT solver~\cite{DBLP:conf/tacas/ClarkeKL04}.

\subsection{Java Operational Model}



The Java language relies on compiler-generated functions and classes as well
as a large standard library.  In order to correctly support Java
functionality, we developed an abstract representation of the standard Java
libraries, called the operational model (OM).  The use of OMs is commonplace
in analyzers for Java; for instance, a similar approach was previously
proposed for the formal verification of Android
applications~\cite{DBLP:journals/sigsoft/MerweTMV15}.  Currently, our OM
consists of models of the most common classes from \texttt{java.lang} and a
few from \texttt{java.util}.%
\footnote{See \url{https://github.com/diffblue/java-models-library}
for the available classes, methods and their limitations.}
Our Java OM simplifies the implementation of
the standard Java library by removing verification-irrelevant performance
optimizations (e.g., in the implementation of container classes), exploiting
declarative specifications (using \emph{assume}) and functions that are
built into the CPROVER framework (e.g., for array and string manipulation).
This built-in functionality is provided through the CPROVER API library.
We are continuously extending our OM to speed up verification by replacing
the original standard Java library classes by our models.

Java has an \texttt{assert($c$)} statement for specifying safety properties. 
In~addition, the CPROVER API provides methods in the
\texttt{org.cprover.CProver} class that allow users to define
non-deterministic verification harnesses and stub functions.  The API
contains such methods for primitive types (e.g.,\ \texttt{int
 nondetInt()}) and \emph{generic} methods (i.e., parametrised by a type
\texttt{T}) as \texttt{<T> T nondetWithNull()} and \texttt{<T> T
nondetWithoutNull()} to non-deterministically initialize object references
that may or may not be \texttt{null}.  The API also provides an
\texttt{assume($c$)} method, which advises JBMC to ignore paths that do not
satisfy a user-specified condition~$c$.

If JBMC encounters an unmodelled library call then it will output a warning
and treat the return value as nondeterministic.%
\footnote{In SV-COMP, the analysis is forced to return \texttt{UNKNOWN}
in such cases.}

CPROVER API and Java library models are available as JAR files that
need to be added to the classpath of the analyzed project, see
Section~\ref{sec:project}.

Currently, JBMC handles neither the Java Native Interface, which
allows Java code to interface native libraries, nor reflection, which
allows the program to inspect and manipulate itself at runtime.  JBMC
has partial support for generics and lambdas as well as multi-threaded
Java programs (that use \texttt{java.lang.Thread}), exploiting the
partial order encoding technique of~\cite{DBLP:conf/cav/AlglaveKT13}.



\subsection{String Constraint Solver}

One of the biggest challenges in verifying Java programs is the widespread
use of character strings, which makes verification problems resulting from
Java programs highly complex.  Solving such constraints is an active area of
research~\cite{DBLP:journals/pacmpl/ChenCHLW18,DBLP:journals/pacmpl/HolikJLRV18,DBLP:journals/fmsd/LiangRTTBD16}.

JBMC implements a solver for strings to determine the satisfiability of a
set of constraints involving string operations.  Our string solver supports
the most common basic accesses (e.g., obtain the length of a string and a
character at a given position); comparisons (e.g., lexicographic comparison
and equality); transformations (e.g., insertion, concatenation, replacement,
and removal); and conversions (e.g., conversion of the primitive data types
into a string and parsing them from a string).

The axioms for these operations involve quantified constraints. For
instance, a Java expression \texttt{s.substring(5)} is translated into
a predicate $\mathit{substring}(\mathit{res}, s, 5)$, where
$\mathit{res}$, $s$ are pairs $(\mathit{length},
\mathit{charArray})$, representing the resulting and the input
string \texttt{s}, respectively; and $\mathit{substring}$ is
axiomatized by the formula $\forall i. (0\le i\,\wedge\,i <
s.\mathit{length}-5)\rightarrow (\mathit{res}.\mathit{length} = s.\mathit{length}-5) \wedge (\mathit{res}.\mathit{charArray}[i]
= s.\mathit{charArray}[i+5])$.  The universal quantifiers are
handled through a quantifier elimination
scheme~\cite{DBLP:conf/hvc/LiG13}.
The string solver implements a
refinement loop, which relies on a set of axioms to check the
satisfiability of the quantified constraints set and an instantiation
function to produce propositional formulas implied by that set of
constraints, which can be solved by a SAT solver.

The string solver is parameterized by a maximum string length
given by the \texttt{{-}-max-nondet-string-length} command line option.
It decides whether a formula involving strings is satisfiable
or unsatisfiable for any string up to that given length.

\subsubsection{Form of the quantified constraints}

Each string is associated with an array of characters and a length,
and the built-in string functions are converted to existential and
universal formulae over arrays. These formulae have a specific form:
$\forall i \in \mathbb{N}.\ \texttt{index\_guard}(i) \rightarrow
\texttt{value\_constraint}(i)$ where:
\texttt{index\_guard} must define an integer
interval, it may contain symbols, but does not reference any string/array;
\texttt{value\_constraint} can access strings at indexes which are linear
functions of $i$ (this is so that it can be reverted).
For instance, $\mathit{insert}(r, s, t, \mathit{offset})$ is encoded by
the formulae:
\begin{eqnarray*}
r.\mathit{length} = s.\mathit{length} + t.\mathit{length} \\
\forall i, 0 \le i < \mathit{offset} \rightarrow r.\mathit{charArray}[i] =
s.\mathit{charArray}[i]\\
\forall i, \mathit{offset} \le i < \mathit{offset} + t.\mathit{length}
\rightarrow r.\mathit{charArray}[i] = t.\mathit{charArray}[i-\mathit{offset}] \\
\forall i, \mathit{offset} + t.\mathit{length} \le i < r.\mathit{length}
\rightarrow r.\mathit{charArray}[i] = s.\mathit{charArray}[i-t.\mathit{length}]
\end{eqnarray*}

\subsubsection{Instantiation}

Each string in the system of equations is associated with a set of indexes
which is progressively expanded in each step of the refinement procedure as
required. Universal formulae are then instantiated using these indexes to obtain
existential formulas corresponding to a more relaxed version of the constraint.
For instance, if the index set of $r$ contains $r.\mathit{length} - 1$, the
formula
$$
\forall i, \mathit{offset} \le i < \mathit{offset} + t.\mathit{length}
\rightarrow r.\mathit{charArray}[i] = t.\mathit{charArray}[i-\mathit{offset}]
$$
will be instantiated as
$$
\begin{array}{r@{}l}
\mathit{offset} &< r.\mathit{length} - 1
< \mathit{offset} + t.\mathit{length} \rightarrow \\
&r.\mathit{charArray}[r.\mathit{length} - 1] = t.\mathit{charArray}[r.\mathit{length} - \mathit{offset}].
\end{array}
$$ The existential formula is added to the system of equations that
are passed to the underlying SAT solver.  At this point, if the underlying
solver answers \emph{unsatisfiable} we can conclude that the original
system is unsatisfiable, since the set of equations we gave to it are
under-constrained.  In the other case, we cannot conclude yet, but
could check if the model given by the solver is actually correct.

\subsubsection{Checking a model}

Assuming the solver answers \emph{satisfiable} for the
under-constrained set of equations, and it can give us a model for the
strings, we can check that the model satisfies the universal
constraints. This is done by taking the negation of the constraint,
which is then an existential one, and substituting symbols by their
values found in the model. This results in a formula where the
original universally quantified variable $i$ is the only unknown, and
its validity can be checked by a call to a SAT solver. If it is valid
then the model found is indeed a model of the original system of
equations. In the other case, the index set needs to be refined.

Note that the model for the strings may be incomplete because the instantiated
formula only refers to them at some specific indexes. In order to make it
complete, we interpret the value at each index by the next one that is known in
the model. In practice, if the solver only says that
$s.\mathit{charArray}[3] = \qchar{a}$ and $s.\mathit{charArray}[5] = \qchar{b}$,
we will replace in the formula $s.\mathit{charArray}[i]$ by
$i \le 3 ? \qchar{a} : \qchar{b}$.

\subsubsection{Refining the index set}

In case we could not conclude from the two above steps above, we add to the
index set by propagating indexes from strings to other strings which appear in
the same equation.
For instance for the formula \(\forall i, \mathit{offset} \le
i < \mathit{offset} + t.\mathit{length}
\rightarrow r.\mathit{charArray}[i] = t.\mathit{charArray}[i-\mathit{offset}] \),
for indexes $j$ that are in the index set
we add the corresponding index $j - \mathit{offset}$ to the index set of $t$.
Reciprocally for indexes $k$ in the index set of $t$ we add
$k + \mathit{offset}$ to the index set of $r$.
The procedure starts back from the instantiation step. It may require several
successive refinements to be able to decide the satisfiability, but when the
fixed point of the index set is reached, one of the two steps above will be
successful.

\section{Evaluations}


JBMC participated~\cite{DBLP:conf/tacas/CordeiroKS19} in the Java track~\cite{DBLP:journals/corr/abs-1809-03739} of SV-COMP'19~\cite{DBLP:conf/tacas/Beyer19} and SV-COMP'20~\cite{DBLP:conf/tacas/Beyer20}, earning the gold and silver medals, respectively. In these SV-COMP editions, we observed some limitations. 

JBMC can only do bounded model checking or full model checking on bounded programs. It has problems with programs that require many unwindings to verify since our BMC engine produces a large formula to be checked by the underlying SAT/SMT solver. 

Our Java operational model is quite incomplete. Many classes that occur in practice are not supported yet. In particular, those classes' methods miss their preconditions and postconditions, and simulation features (e.g., how elements are manipulated). As a result, JBMC over-approximates the program's original semantics.

JBMC's support for multi-threading is not very efficient. Although JBMC models program executions with partial orders rather than interleavings, it still makes automatic analysis of multi-threaded Java code very hard.

Our string solver currently does not support regular expressions. There is no support for reflection or native methods.


%
%

\section{Conclusions and Future Directions}

JBMC is a bounded model checker for Java Bytecode. It builds upon the CPROVER framework and shares a large part of the backend functionality with CBMC. Although JBMC is a relatively recent tool, it is currently one of the best model checkers for Java if we consider SV-COMP's latest editions~\cite{DBLP:conf/tacas/Beyer19,DBLP:conf/tacas/Beyer20}. However, it still has many limitations regarding the features of the Java language that it supports.


\begin{thebibliography}{10}
\providecommand{\url}[1]{{#1}}
\providecommand{\urlprefix}{URL }
\expandafter\ifx\csname urlstyle\endcsname\relax
  \providecommand{\doi}[1]{DOI~\discretionary{}{}{}#1}\else
  \providecommand{\doi}{DOI~\discretionary{}{}{}\begingroup
  \urlstyle{rm}\Url}\fi

\bibitem{Alexander02}
Alexander, R.T., Offutt, J., Bieman, J.M.: Fault detection capabilities of
  coupling-based {OO} testing.
\newblock In: Software Reliability Engineering, pp. 207--2002 (2002)

\bibitem{DBLP:conf/cav/AlglaveKT13}
Alglave, J., Kroening, D., Tautschnig, M.: Partial orders for efficient bounded
  model checking of concurrent software.
\newblock In: {CAV}, \emph{LNCS}, vol. 8044, pp. 141--157 (2013)

\bibitem{DBLP:conf/tacas/AnandPV07}
Anand, S., Pasareanu, C.S., Visser, W.: {JPF-SE:} {A} symbolic execution
  extension to {Java} {PathFinder}.
\newblock In: {TACAS}, \emph{LNCS}, vol. 4424, pp. 134--138 (2007)

\bibitem{BarrettSST09}
Barrett, C., Sebastiani, R., Seshia, S.A., Tinelli, C.: Satisfiability Modulo
  Theories, \emph{Frontiers in Artificial Intelligence and Applications}, vol.
  185, chap.~26, pp. 825--885.
\newblock IOS Press (2009)

\bibitem{Barrett10c.:the}
Barrett, C., Stump, A., Tinelli, C., Boehme, S., Cok, D., Deharbe, D.,
  Dutertre, B., Fontaine, P., Ganesh, V., Griggio, A., Grundy, J., Jackson, P.,
  Oliveras, A., Krstić, S., Moskal, M., De~Moura, L., Sebastiani, R., Cok,
  T.D., Hoenicke, J.: {T}he {SMT}-{LIB} {S}tandard: {V}ersion 2.0.
\newblock Tech. rep. (2010)

\bibitem{DBLP:conf/tacas/Beyer19}
Beyer, D.: Automatic verification of {C} and java programs: {SV-COMP} 2019.
\newblock In: Tools and Algorithms for the Construction and Analysis of
  Systems, \emph{Lecture Notes in Computer Science}, vol. 11429, pp. 133--155
  (2019).
\newblock \doi{10.1007/978-3-030-17502-3\_9}

\bibitem{DBLP:conf/tacas/Beyer20}
Beyer, D.: Advances in automatic software verification: {SV-COMP} 2020.
\newblock In: TACAS, \emph{LNCS}, vol. 12079, pp. 347--367 (2020).
\newblock \doi{10.1007/978-3-030-45237-7\_21}

\bibitem{handbook09}
Biere, A., Heule, M., van Maaren, H., Walsh, T. (eds.): Handbook of
  Satisfiability, \emph{Frontiers in Artificial Intelligence and Applications},
  vol. 185. {IOS} Press (2009)

\bibitem{DBLP:journals/pacmpl/ChenCHLW18}
Chen, T., Chen, Y., Hague, M., Lin, A.W., Wu, Z.: What is decidable about
  string constraints with the {ReplaceAll} function.
\newblock {PACMPL} \textbf{2}({POPL}), 3:1--3:29 (2018)

\bibitem{ChudnovCCDHMMMM18}
Chudnov, A., Collins, N., Cook, B., Dodds, J., Huffman, B., MacC{\'{a}}rthaigh,
  C., Magill, S., Mertens, E., Mullen, E., Tasiran, S., Tomb, A., Westbrook,
  E.: Continuous formal verification of amazon s2n.
\newblock In: {CAV}, \emph{LNCS}, vol. 10982, pp. 430--446. Springer (2018)

\bibitem{CKL04}
Clarke, E., Kroening, D., Lerda, F.: A tool for checking {ANSI-C} programs.
\newblock In: Tools and Algorithms for the Construction and Analysis of
  Systems, \emph{Lecture Notes in Computer Science}, vol. 2988, pp. 168--176.
  Springer (2004)

\bibitem{DBLP:conf/tacas/ClarkeKL04}
Clarke, E.M., Kroening, D., Lerda, F.: A tool for checking {ANSI-C} programs.
\newblock In: {TACAS}, \emph{LNCS}, vol. 2988, pp. 168--176 (2004)

\bibitem{CordeiroFCM09}
Cordeiro, L.C., Fischer, B., Chen, H., Marques{-}Silva, J.: Semiformal
  verification of embedded software in medical devices considering stringent
  hardware constraints.
\newblock In: {ICESS}, pp. 396--403. {IEEE} Computer Society (2009)

\bibitem{CKKST18}
Cordeiro, L.C., Kesseli, P., Kroening, D., Schrammel, P., Trt{\'{\i}}k, M.:
  {JBMC:} {A} bounded model checking tool for verifying {Java} bytecode.
\newblock In: Computer Aided Verification, {CAV}, \emph{LNCS}, vol. 10981, pp.
  183--190. Springer (2018)

\bibitem{DBLP:journals/corr/abs-1809-03739}
Cordeiro, L.C., Kroening, D., Schrammel, P.: Benchmarking of java verification
  tools at the software verification competition {(SV-COMP)}.
\newblock CoRR \textbf{abs/1809.03739} (2018).
\newblock \urlprefix\url{http://arxiv.org/abs/1809.03739}

\bibitem{DBLP:conf/tacas/CordeiroKS19}
Cordeiro, L.C., Kroening, D., Schrammel, P.: {JBMC:} bounded model checking for
  java bytecode - (competition contribution).
\newblock In: Tools and Algorithms for the Construction and Analysis of
  Systems, \emph{Lecture Notes in Computer Science}, vol. 11429, pp. 219--223
  (2019).
\newblock \doi{10.1007/978-3-030-17502-3\_17}.
\newblock \urlprefix\url{https://doi.org/10.1007/978-3-030-17502-3\_17}

\bibitem{Deitel2019}
Deitel, H.M., Deitel, P.J.: {Java} How to Program, Late Objects, global edition
  edn.
\newblock Prentice Hall Press (2019)

\bibitem{DBLP:journals/sigplan/FlanaganLLNSS13}
Flanagan, C., Leino, K.R.M., Lillibridge, M., Nelson, G., Saxe, J.B., Stata,
  R.: {PLDI} 2002: Extended static checking for {Java}.
\newblock {SIGPLAN} Notices \textbf{48}(4S), 22--33 (2013)

\bibitem{DBLP:journals/pacmpl/HolikJLRV18}
Hol{\'{\i}}k, L., Janku, P., Lin, A.W., R{\"{u}}mmer, P., Vojnar, T.: String
  constraints with concatenation and transducers solved efficiently.
\newblock {PACMPL} \textbf{2}({POPL}), 4:1--4:32 (2018)

\bibitem{DBLP:conf/cav/KahsaiRSS16}
Kahsai, T., R{\"{u}}mmer, P., Sanchez, H., Sch{\"{a}}f, M.: {JayHorn}: {A}
  framework for verifying {Java} programs.
\newblock In: {CAV}, \emph{LNCS}, vol. 9779 (2016)

\bibitem{DBLP:conf/hvc/LiG13}
Li, G., Ghosh, I.: {PASS:} string solving with parameterized array and interval
  automaton.
\newblock In: HVC, \emph{LNCS}, vol. 8244, pp. 15--31 (2013)

\bibitem{DBLP:journals/fmsd/LiangRTTBD16}
Liang, T., Reynolds, A., Tsiskaridze, N., Tinelli, C., Barrett, C., Deters, M.:
  An efficient {SMT} solver for string constraints.
\newblock Formal Methods in System Design \textbf{48}(3), 206--234 (2016)

\bibitem{DBLP:journals/sigsoft/MerweTMV15}
van~der Merwe, H., Tkachuk, O., van~der Merwe, B., Visser, W.: Generation of
  library models for verification of {Android} applications.
\newblock {ACM} {SIGSOFT} Software Engineering Notes \textbf{40}(1), 1--5
  (2015)

\bibitem{Muchnick97}
Muchnick, S.S.: Advanced compiler design and implementation.
\newblock Morgan Kaufmann Publishers Inc. (1997)

\bibitem{OHearn18}
O'Hearn, P.W.: Continuous reasoning: Scaling the impact of formal methods.
\newblock In: {LICS}, pp. 13--25. {ACM} (2018)

\bibitem{javadoc2017}
Oracle: Java$^{TM}$ programming language.
\newblock
  \url{https://docs.oracle.com/javase/8/docs/technotes/guides/language/index.html}
  (2017).
\newblock Accessed: 31-01-2018

\end{thebibliography}

\end{document}